\documentstyle[psfig]{cici}

\def\simg{\mathrel{%
      \rlap{\raise 0.511ex \hbox{$>$}}{\lower 0.511ex \hbox{$\sim$}}}}
\def\siml{\mathrel{%
      \rlap{\raise 0.511ex \hbox{$<$}}{\lower 0.511ex \hbox{$\sim$}}}}
\def\eq{eq$.$} \def\eqs{eqs$.$} \def\etal{et al$.$ } \def\eg{e.g$.$ } \def\ie{i.e$.$ }
\def\h{\hspace*{-3mm}} \def\ds{\displaystyle} \def\p{$\pm$} \def\d{\rm d}  
\def\S{{\it S}~} \def\j{{\it j}~} \def\J{{\it J}~} 
\def\Sa{{\it Sa}~} \def\ja{{\it ja}~} \def\Ja{{\it Ja}~}
\def\Sc{{\it Sc}~} \def\jc{{\it jc}~} \def\Jc{{\it Jc}~}
\def\RSc{{\it RSc}~} \def\Rjc{{\it Rjc}~} 
\def\reference{\bibitem}

\begin{document}

\title [ Nine Swift $X$-ray Afterglows ]
  { Analysis of the $X$-ray Emission of Nine Swift Afterglows }

\author[Panaitescu \etal ]{ A. Panaitescu$^1$, P. M\'esz\'aros$^{2,3}$, N. Gehrels$^4$, D. Burrows$^2$, 
         J. Nousek$^2$ \\ 
$^1$ Space Science and Applications, MS D466, Los Alamos National Laboratory, Los Alamos, NM 87545 \\
$^2$ Department of Astronomy \& Astrophysics, Pennsylvania State University, University Park, PA 16802 \\ 
$^3$ Department of Physics, Pennsylvania State University, University Park, PA 16802 \\ 
$^4$ NASA/Goddard Space Flight Center, Greenbelt, MD 20771 }

\maketitle

\begin{abstract} 
\begin{small}
  The $X$-ray light-curves of 9 Swift XRT afterglows (050126, 050128, 050219A, 050315, 050318,
  050319, 050401, 050408, 050505) display a complex behaviour: a steep $t^{-3.0 \pm 0.3}$ 
  decay until $\sim 400$ s, followed by a significantly slower $t^{-0.65 \pm 0.20}$ fall-off,
  which at 0.2--2 d after the burst evolves into a $t^{-1.7 \pm 0.5}$ decay. 
  We consider three possible models for the geometry of relativistic blast-waves (spherical 
  outflows, non-spreading jets, and spreading jets), two possible dynamical regimes for the
  forward shock (adiabatic and fully radiative), and we take into account a possible angular
  structure of the outflow and delayed energy injection in the blast-wave, to identify the 
  models which reconcile the $X$-ray light-curve decay with the slope of the $X$-ray continuum 
  for each of the above three afterglow phases. 
  By piecing together the various models for each phase in a way that makes physical sense, 
  we identify possible models for the entire $X$-ray afterglow. 
  The major conclusion of this work is that a long-lived episode of energy injection in the 
  blast-wave, during which the shock energy increases at $t^{1.0\pm0.5}$, is required for 5 
  afterglows and could be at work in the other 4 as well.
  For some afterglows, there may be other mechanisms that can explain the $t < 400$ s fast 
  falling-off $X$-ray light-curve (\eg the large-angle GRB emission), the 400 s--5 h slow
  decay (\eg a structured outflow), or the steepening at 0.2--2d (\eg a jet-break, a 
  collimated outflow transiting from a wind with a $r^{-3}$ radial density profile to a 
  homogeneous or outward-increasing density region). 
  Optical observations in conjunction with the $X$-ray can distinguish among these various models. 
  Our simple tests allow the determination of the location of the cooling frequency relative 
  to the $X$-ray domain and, thus, of the index of the electron power-law distribution with 
  energy in the blast-wave. The resulting indices are clearly inconsistent with an universal value.
\end{small}  
\end{abstract}


\section{Introduction}

 Pre-Swift observations of Gamma-Ray Burst (GRB) afterglows have led to great strides in their
theoretical interpretation, while leaving some major unanswered questions.

 The radio, optical, and $X$-ray emission of GRB afterglows exhibit a power-law decrease with time 
($F_\nu \propto t^{-\alpha}$) from hours to tens of days after the burst, with the temporal index 
$\alpha$ consistent with the slope $\beta$ of the power-law continuum ($F_\nu \propto \nu^{-\beta}$) 
within the framework of relativistic spherical blast-waves (M\'esz\'aros \& Rees 1997) or of spreading 
relativistic jets (Rhoads 1999). The collimation of the GRB ejecta yields a steepening of the power-law 
decay when the relativistic beaming has decreased sufficiently that the jet boundary becomes visible. 
Such a steepening has been observed for the first time in the optical light-curve of the afterglow 990123 
(Kulkarni \etal 1999). Since then about 10 other afterglows have displayed an optical light-curve break 
at about 1 day after the burst. The achromaticity of a jet light-curve break has not been clearly proven 
by pre-Swift observations because the $X$-ray light-curves were not monitored over a time long enough to 
capture the jet-break. Furthermore, the radio light-curves were usually poorly sampled during the first
day and strongly affected by interstellar scintillation. The observations of many $X$-ray afterglows by 
Swift, together with ground-based optical observations, will enable us to test achromaticity of the 
afterglow light-curve break, as appears to be the case for the afterglow 050525A (Blustin \etal 2005).

 The quenching of the interstellar scintillation of the radio afterglow 970508 (Frail, Waxman \& Kulkarni 
2000) has confirmed that the source size increases as expected for a relativistic blast-wave, providing 
another test for this model. The decrease of the scintillation has also been observed in the radio 
afterglows 991208, 021004, and 030329.
Further testing has been prompted by the detection of the optical afterglows of GRBs 990123 and 021211 
at very early times (Akerlof \etal 1999, Fox \etal 2003, Li \etal 2003), starting at about 100 s after 
the burst. The steep decays ($\alpha = 1.8$ and 1.6, respectively) exhibited by these afterglows in the 
first 20 minutes can be attributed to the GRB ejecta energized by internal shocks (M\'esz\'aros \& Rees 
1997, 1999) or by the reverse shock\footnotemark caused by the interaction with the circumburst medium 
(Sari \& Piran 1999). 
\footnotetext{ Later it became clear that the latter interpretation is not so straightforward: 
    the reverse-shock microphysical parameters required to accommodate the early optical light-curves 
    of the afterglows 990123 and 021211 imply a magnetized ejecta (Fan \etal 2002, Kumar \& Panaitescu 
    2003, Zhang, Kobayashi \& M\'esz\'aros 2003) }

 A significant discrepancy between afterglow observations and theoretical expectations exists for the
radio afterglows of GRBs 991208, 991216, 000301c, and 010222, whose decay over 1--2 decades in time is 
substantially slower than that of the optical emission (Frail \etal 2004, Panaitescu \& Kumar 2004). 
A change in the blast-waves dynamics, such as the transition to semi-relativistic dynamics, is not a 
possible explanation, because the different radio and optical decays are observed over time ranges which 
overlap substantially. Our analysis (Panaitescu \& Kumar 2004) of these afterglows shows that evolving 
microphysical parameters cannot decouple the optical and radio decays. This decoupling may be achieved 
if there is an extra radio emission arising from some late ejecta, energized by a reverse shock. 
For the optical afterglow to remain unaffected, the incoming ejecta should not alter the dynamics of 
the blast-wave, \ie they should carry less kinetic than that already existing in the swept-up circumburst 
medium.

 The Swift measurements of the $X$-ray afterglow emission, starting from 100 s after the burst, opens 
new possibilities for testing the blast-wave model and for refining its details. The XRT 0.2--10 keV 
light-curves of the 9 $X$-ray afterglows (050126, 050128, 050219A, 050315, 050318, 050319, 050401, 050408, 
050505) presented by Campana \etal (2005), Chincarini \etal (2005), and Tagliaferri \etal (2005), have
shown that some $X$-ray afterglows decay very fast ($F_x \propto t^{-3}$) within the first few minutes 
after the burst, as reported previously for the afterglows 990510 (Pian \etal 2001) and 010222 (in't Zand 
\etal 2001), followed by a slower decay phase ($F_x \propto t^{-2/3}$), and a break to a steeper decay 
($F_x \propto t^{-5/3}$) at a later time, ranging from 1 hour to 1 day. The purpose of this paper is to 
investigate what features of the blast-wave model are required to accommodate the various decays of these 
Swift $X$-ray afterglows. 

 Barthelmy \etal (2005) have shown that the very early fast decay of the $X$-ray emission of the afterglows 
050315 and 050319 can be understood as the GRB emission from the fluid moving at angles larger than the 
inverse of the forward shock's Lorentz factor. Due to the curvature of the emitting surface, this large
angle emission arrives at the observer at an ever increasing time and ever decreasing frequency (Kumar \& 
Panaitescu 2000). However, for two other Swift afterglows with an early, fast decaying $X$-ray emission 
(050126 and 050219A), Tagliaferri \etal (2005) have found that the 15--350 keV GRB emission extrapolated 
to the XRT 0.2--10 keV band (under the assumption that the burst power-law spectrum extends unbroken to
lower energies) falls short of the flux measured at the beginning of the $X$-ray observations. If the
burst spectrum has a break below 15 keV, below which it is harder, then the GRB extrapolated flux would be 
even less. This suggests that the fast falling-off $X$-ray emission does not arise from the same mechanism
as the burst itself, and that it may arise in the forward shock. Therefore, for at least these two last bursts, 
we shall test whether the very early $X$-ray emission can have the same origin as the rest of the afterglow. 

 The steepening observed at later times is most naturally attributed to a collimated outflow (jet), hence we 
shall test if the pre- and post-break $X$-ray light-curve indices and the spectral slopes are consistent with 
this interpretation.

\section{The $X$-ray Light-Curve Decay Index}
\label{theory}
 
 For the dynamics and collimation of the relativistic blast-wave, we consider three cases:
$i)$ a spherical GRB remnant, in the sense that observations were done at a time when the 
afterglow Lorentz factor $\Gamma$ was larger than the inverse of the jet opening $\theta_j$ 
and, hence, the effects associated with collimation were not yet detectable, 
$ii)$ a jet whose edge is visible ($\Gamma \theta_j < 1$) and which does not expand laterally 
(because it is embedded in an outer outflow, but whose emission is dimmer), and 
$iii)$ a jet with sharp edges, which spreads laterally and is observed when $\Gamma \theta_j < 1$.
These models will be named \S, \j, and \J, respectively.

 At a frequency above that of the synchrotron peak, $\nu_i$, the index $\alpha$ of the light-curve 
power-law decay depends on \\
$i)$ the index $p$ of the power-law electron distribution with energy 
\begin{equation}
 \frac{\d N}{d\epsilon} \propto \epsilon^{-p} \;, 
\label{N}
\end{equation}
$ii)$ the density stratification of the circumburst medium (CBM), for which we assume a power-law 
profile 
\begin{equation}
 n(r) \propto r^{-s} \quad s < 3 \;,
\label{n}
\end{equation}
which comprises a homogeneous CBM ($s=0$) and a pre-ejected wind at constant speed and mass-loss rate 
($s=2$), the condition $s < 3$ being required for a decelerating blast-wave, \\
$iii)$ the location of the cooling frequency $\nu_c$ relative to the observing band. The $\nu_c$
      is the synchrotron characteristic frequency corresponding to an electron energy for which
      the radiative (synchrotron + inverse Compton) timescale is equal to the electron age. \\
The expressions for $\alpha(p,s)$ for the \S model are 
given in M\'esz\'aros \& Rees (1997) and Sari, Narayan \& Piran (1998) for $s=0$ and in Chevalier 
\& Li (2000) for $s=2$. Rhoads (1999) and Sari, Piran \& Halpern (1999) have shown that, for the 
\J model, $\alpha = p$, irrespective of the location of $\nu_c$ and CBM stratification. 
These and other results for the \S and \j models are summarized below.

\subsection{Adiabatic Afterglows}

 Because we will determine from observations the required structure of the CBM, \ie the parameter
$s$, we start from the most general expressions for the evolution with observer time $t$ of the 
afterglow spectral properties: peak flux $F_p$, and frequencies $\nu_i$ and $\nu_c$. As derived 
in Panaitescu \& Kumar (2004), they are:
\begin{equation}
 (S,j)\quad  \frac{\ds \d\ln\nu_i}{\ds \d\ln t} = -\frac{\ds 3}{\ds 2} \;,\quad 
               \frac{\ds \d\ln\nu_c}{\ds \d\ln t} = \frac{\ds 3s-4}{\ds 8-2s} 
\label{nuic}
\end{equation}
for both the \S and \j models (note that the evolution of $\nu_i$ is in independent of $s$
and that $\nu_c$ increases for $s < 4/3$, but decreases for $s > 4/3$),
\begin{equation}
 (S) \quad \frac{\ds \d\ln F_p}{\ds \d\ln t} = -\frac{\ds s}{\ds 8-2s} 
\end{equation}
for the \S model and
\begin{equation}
 (j) \quad \frac{\ds \d\ln F_p}{\ds \d\ln t} = -\frac{\ds 6-s}{\ds 8-2s} 
\label{Fpj}
\end{equation}
for the \j model. For the latter, the faster decay is due to that the jet area is a factor 
$(\Gamma \theta_j)^2$ smaller than that visible to the observer in the case of a spherical
outflow.

 The synchrotron afterglow continuum is $F_\nu \propto \nu^{-\beta}$ (Sari, Narayan \& Piran 1998), 
where
\begin{equation}
 \beta = \left\{ \begin{array}{ll}  1/2     & \nu_c < \nu < \nu_i  \\
                                    (p-1)/2 & \nu_i < \nu < \nu_c  \\
                                    p/2     & \nu_i,\nu_c < \nu    \end{array} \right. \;.
\label{beta}
\end{equation}
We restrict our attention to the $\nu > \min\{\nu_i,\nu_c\}$ cases, for which $\beta > 1/2$,
as observed by XRT for the Swift $X$-ray afterglows. From equations (\ref{nuic})--(\ref{Fpj}), 
it is easy to obtain the synchrotron light-curve decay $F_\nu \propto t^{-\alpha}$:
\begin{equation}
 (Sa)\quad \alpha (\nu_i < \nu < \nu_c) - \frac{\ds 3}{\ds 2}\beta = \frac{\ds s}{\ds 8-2s} =
   \left\{ \begin{array}{ll} -1/2 \,&\, s \rightarrow -\infty \\ 0 \,&\, s=0 \\ 
                              1/2 \,&\, s=2 \\ 3/2 \,&\, s=3 \end{array} \right. 
\end{equation}
\label{Sa}
\begin{equation}
 (Sc)\quad \alpha (\nu > \nu_c) - \frac{\ds 3}{\ds 2}\beta = -\frac{\ds 1}{\ds 2} 
\label{Sc}
\end{equation}
\begin{equation}
 (ja)\quad \alpha (\nu_i < \nu < \nu_c) - \frac{\ds 3}{\ds 2}\beta = \frac{\ds 6-s}{\ds 8-2s} =
   \left\{ \begin{array}{ll} 1/2 \,&\, s \rightarrow -\infty \\ 3/4 \,&\, s=0 \\ 
                               1 \,&\, s=2 \\ 3/2 \,&\, s=3 \end{array} \right. 
\label{ja}
\end{equation}
\begin{equation}
 (jc)\quad \alpha (\nu > \nu_c) - \frac{\ds 3}{\ds 2}\beta = \frac{\ds 2-s}{\ds 8-2s} =
   \left\{ \begin{array}{ll} 1/2 \,&\, s \rightarrow -\infty \\ 1/4 \,&\, s=0 \\ 
                               0 \,&\, s=2 \\ -1/2 \,&\, s=3 \end{array} \right. \;.
\label{jc}
\end{equation}

 From the above equations, it can be seen that the passage of the cooling frequency through the
observing band steepens the afterglow decay by $\Delta \alpha = |4-3s|/(16-4s)$, which is at most
1/4,in addition to softening the spectrum  by $\Delta \beta = 1/2$ for $s < 4/3$ or hardening
it by $\Delta \beta = -1/2$ for $s > 4/3$. The representative values chosen for $s$ these equations 
show that the observable quantity $\alpha - 1.5 \beta$ has a stronger dependence on the CBM 
structure for $s \siml 3$ (winds) than for $s \sim 0$. The case $s=3$ should be taken only as
the $s \rightarrow 3$ limit; for $s=3$ the outflow deceleration is not a power-law in the 
observer time, instead $\Gamma \propto 1/\sqrt{\ln t}$.

 For the \J model, the ($\alpha,\beta$) closure relation is:
\begin{equation}
 (Ja)\quad \alpha (\nu_i < \nu < \nu_c) - 2\beta = 1 
\label{Ja}
\end{equation}
\begin{equation}
 (Jc)\quad \alpha (\nu > \nu_c) - 2\beta = 0 \;. 
\label{Jc}
\end{equation}

 Equations (\ref{Sc}), (\ref{jc}), and (\ref{Jc}) are valid whatever is the location of the 
injection frequency. However, there are further constraints for the applicability of the 
$\nu_c < \nu < \nu_i$ case: $\beta = 1/2$ for all models and $\alpha = 1/4,1$ for the \S and \J 
models respectively. 

 The models \S, \j, and \J with $\nu < \nu_c$ will be designated as \Sa, \ja, and \Ja (the letter 
"a" indicating that the electrons radiating synchrotron emission at the observing frequency are 
losing energy adiabatically), while the models with $\nu_c < \nu$ will be called \Sc, \jc, and \Jc 
(where the letter "c" shows that the electrons radiating at $\nu$ are cooling radiatively). 
Note from equations (\ref{Sc}), (\ref{Ja}), and (\ref{Jc}) that for the \Sc and \J models, 
the index $\alpha$ is independent of the medium structure, hence the type of CBM cannot be 
determined for these models.

\subsection{Inverse-Compton Dominated Electron Cooling}

 In the derivation of equation (\ref{nuic}) we have ignored a multiplicative factor 
$(Y+1)^{-2}$ (where $Y$ is the Compton parameter) in the expression of $\nu_c$. Therefore 
equation (\ref{nuic}) is valid if $Y < 1$ (\ie the radiative cooling of the electrons emitting 
at $\nu_c$ is synchrotron-dominated) or if $Y$ constant (which corresponds to the $\nu_c < \nu_i$ 
case, where the $Y$ 
parameter depends only on the ratio of the electron and magnetic field energies). If $Y > 1$ and 
$\nu_i < \nu_c$, the decrease of the Compton parameter with time leads to a faster increase 
or a slower decrease of $\nu_c$ than given in equation (\ref{nuic}) and to a slower decay of 
the afterglow emission at $\nu > \nu_c$. This case is most likely relevant for the $2^{nd}$ 
$X$-ray afterglow phase, between the flattening and steepening times $t_F$ and $t_S$, when the 
$X$-ray light-curve may exhibit a slower decay than that resulting from equations (\ref{Sc}), 
(\ref{jc}), and (\ref{Jc}). The equations for the afterglow light-curve at $\nu > \nu_c$ for
the ($Y > 1$, $\nu_i < \nu_c$) case, derived by Panaitescu \& Kumar (2001), lead to:  \\ \\
\begin{equation}
 (Sc,s=0)\quad \alpha - \frac{\ds 3}{\ds 2}\beta = \left\{ 
         \begin{array}{ll} -1/(4-2\beta) &  p < 3 \\ -1 & p > 3 \end{array}\right. 
\label{Sc0}
\end{equation}
\begin{equation}
 (Sc,s=2)\quad \alpha - \frac{\ds 3}{\ds 2}\beta = \left\{ 
         \begin{array}{ll} -\beta/(4-2\beta) &  p < 3 \\ -3/2 & p > 3 \end{array}\right. 
\label{Sc2}
\end{equation}
\begin{equation}
 (jc,s=0)\quad \alpha - \frac{\ds 3}{\ds 2}\beta = \left\{ \begin{array}{ll} 
            (4-3\beta)/(8-4\beta) &  p < 3 \\ -1/4 & p > 3 \end{array}\right. 
\label{jc0}
\end{equation}
\begin{equation}
 (jc,s=2)\quad \alpha - \frac{\ds 3}{\ds 2}\beta = \left\{ \begin{array}{ll} 
            (1-\beta)/(2-\beta) &  p < 3 \\ -1 & p > 3 \end{array}\right.  \;.
\label{jc2}
\end{equation}
For simplicity, the results in equations (\ref{Sc0})--(\ref{jc2}) are given for the two
most likely types of CBM structure -- $s=0$ and $s=2$ -- and not for any $s$.
For the \J model, $\alpha$ is quasi-independent on the stratification of the CBM:
\begin{equation}
 (Jc)\quad \alpha - 2\beta = \left\{ \begin{array}{ll} 
            (1-\beta)/(2-\beta) & p < 3 \\ -1 & p > 3 \end{array}\right. \;.
\label{JcY}
\end{equation}

 The results given for $p < 3$ in equations (\ref{Sc0})--(\ref{JcY}) are valid also for
$p < 2$ as long as the total electron energy is a constant fraction of the post-shock
energy, which is equivalent to saying that the high-energy cut-off of the electron distribution,
which must exist for a finite total electron energy, has the same evolution as the minimum
electron energy, $\gamma_i \propto \Gamma$.

 The above equations show that the passage of the cooling frequency through the observing 
domain slows the afterglow decay by $\Delta \alpha \geq -1/4$ for $s=0$ and $\Delta \alpha 
\geq -5/4$.

\subsection{Radiative Afterglows}
\label{radiative}

 The temporal evolutions given in equations (\ref{nuic}) and (\ref{Fpj}) were derived under the 
assumption of an adiabatic blast-wave. If the electron fractional energy is around 50 percent and 
if the electrons cool radiatively ($\nu_c < \nu_i$), then radiative losses become important.
In this case the afterglow emission decays faster than for an adiabatic GRB remnant, given 
the stronger deceleration, therefore radiative blast-waves should be of importance for the
fast decaying, very early Swift $X$-ray afterglows.

 From $i)$ the dynamics of a fully radiative blast-wave ($\Gamma M = const$, where
$M \propto nR^3 \propto R^{3-s}$ is the mass of the swept-up CBM) and using $ii)$ the scalings 
for the spectral characteristics ($\nu_{i,c} \propto \gamma_{i,c}^2 B \Gamma$, where
$B \propto \Gamma n^{1/2}$ is the post-shock magnetic field strength, $\gamma_i \propto \Gamma$ 
is the electron energy, and $\gamma_c \propto \Gamma^3 r^{-2} B^{-3}$ is the energy of the
electrons whose radiative cooling timescale is equal to the dynamical timescale; 
$F_p \propto \Gamma B M$ for the \S model and $F_p \propto \Gamma^3 B M$ for the \j model),
and $iii)$ the relation between the observer time and blast-wave radius $r \propto \Gamma^2 t$,
the following evolutions of the spectral characteristics can be derived:
\begin{equation}
 (RS,Rj)\quad  \frac{\ds \d\ln\nu_i}{\ds \d\ln t} = -\frac{\ds 24-7s}{\ds 14-4s} \;,\quad
               \frac{\ds \d\ln\nu_c}{\ds \d\ln t} = \frac{\ds 3s-4}{\ds 14-4s}
\label{Rnuic}
\end{equation}
\begin{equation}
 (RS) \quad \frac{\ds \d\ln F_p}{\ds \d\ln t} = -\frac{\ds 6-s}{\ds 14-4s}
\end{equation}
\begin{equation}
 (Rj) \quad \frac{\ds \d\ln F_p}{\ds \d\ln t} = -\frac{\ds 18-5s}{\ds 14-4s} \;.
\label{RFpj}
\end{equation}
Hereafter, radiative afterglows will be indicated with the letter "R" preceding the specific
model. Note from equation (\ref{Rnuic}) that, just as for an adiabatic afterglow, the cooling
frequency increases for $s < 4/3$ and decreases for $s > 4/3$.

 The light-curve decay indices resulting from equations (\ref{beta}) and (\ref{Rnuic})--(\ref{RFpj}) 
are:
\begin{equation}
 (RSc)\quad \alpha(\nu_c < \nu < \nu_i) = \frac{\ds 16-5s}{\ds 28-8s} 
\end{equation} 
\begin{equation}
 (RSc)\quad \alpha(\nu_c < \nu_i < \nu) = \frac{\ds 24\beta-4-s(7\beta-1)}{\ds 14-4s} 
\end{equation} 
\begin{equation}
 (Rjc)\quad \alpha(\nu_c < \nu < \nu_i) = \frac{\ds 40-13s}{\ds 28-8s} 
\end{equation} 
\begin{equation}
 (Rjc)\quad \alpha(\nu_c < \nu_i < \nu) = \frac{\ds 24\beta+8-s(7\beta+3)}{\ds 14-4s}  \;.
\end{equation} 

 The condition $\nu_c < \nu_i$ required by radiative dynamics guarantees that the Compton 
parameter $Y$ is constant, hence there are no further complications with the inverse
Compton-dominated electron cooling, as it was the case for an adiabatic blast-wave.
Given that, in the \J model, the jet Lorentz factor decreases exponentially with radius
(Rhoads 1999), the dynamics and light-curves of a radiative jet should be close to those
for an adiabatic jet (\eqs [\ref{Ja}] and [\ref{Jc}]).

\subsection{Structured Outflows}
\label{SO}

 There are two other factors which can alter the afterglow decay index $\alpha$. One is that 
the relativistic outflow can be endowed with an angular structure, where the ejecta kinetic 
energy per solid angle, $\d E/\d\Omega$, is not constant (M\'esz\'aros, Rees \& Wijers 1998). 
The light-curve decay indices for an axially-symmetric outflow with a power-law structure 
\begin{equation}
 \frac{\d E}{\d\Omega} \propto \theta^q \;, 
\label{Etheta}
\end{equation}
where the angle $\theta$ is measured from the symmetry axis (which, for simplicity, is assumed 
to be also the direction toward the observer) are given in M\'esz\'aros, Rees \& Wijers (1998) 
and Panaitescu \& Kumar (2003). 
In this work, recourse to a structured outflow will be made only to explain afterglow decays
which are slower than that expected for the \S model. Evidently, such structured outflows 
require $q > 0$. If the slow $X$-ray decay is preceded by a faster fall-off, then the index 
$q$ changes to $q < 0$ close to the outflow axis, corresponding to the \j or \J models.
If the slow $X$-ray decay is followed by a steepening, then, going away from the outflow axis, 
the index $q$ changes to either $q=0$ (if the steeper decay is accommodated by the \S model) 
or to $q < 0$ (if that steeper decay can be explained with the \j and \J models). 
Therefore, in the most general case, where the $X$-ray light-curve exhibits a sharp decay
followed by a slow fall-off and then a steeper dimming, the outflow should have a bright spot 
moving toward the observer, surrounded by a dim envelope (so that a steep decay is obtained 
when the spot edge becomes visible to the observer), which is embedded in a more energetic 
outer outflow (yielding the slower decay), whose collimation leads to the late steepening when 
the outflow boundary becomes visible.

 The decay index for the synchrotron emission from a structured outflow can be derived as described 
in Panaitescu \& Kumar (2003). For a power-law radial structure of the CBM and angular structure 
of the outflow, we obtain:
\begin{equation}
  \alpha (\nu_i < \nu < \nu_c) - \frac{\ds 3}{\ds 2}\beta = 
        \frac{ \ds s - 0.5 q [(3-s)\beta + 6 - 3s] }{ \ds 8 - 2s + q }
\label{SOa}
\end{equation}
\begin{equation}
  \alpha (\nu > \nu_i,\nu_c) - \frac{\ds 3}{\ds 2}\beta = 
        - \frac{\ds 4-s + 0.5 q [(3-s) \beta + 4-s] } { \ds 8 - 2s + q }  \;.
\label{SOc}
\end{equation}
The above results are valid for
\begin{equation}
 q > \tilde{q} \equiv -2(4-s) \cdot \left\{ \begin{array}{ll} 
            [ 3 + 2\beta - 0.5 s (\beta + 3) ]^{-1} & \nu_i < \nu < \nu_c \\
       \left[ 3 + 2\beta - 0.5 s (\beta + 1) \right]^{-1} & \nu_i,\nu_c < \nu 
               \end{array} \right.   \;,
\end{equation}
because for $q < \tilde{q} < 0$ the emission from the outflow axis ($\theta = 0$), where the energy 
per solid angle would formally diverge, becomes dominant and sets another light-curve decay index.
From equations (\ref{SOa}) and (\ref{SOc}) it follows that, for a given CBM structure, the slowest 
decay that a structured outflow can produce is that obtained in the $q \rightarrow \infty$ limit:
\begin{equation}
 \alpha_{min} = \left\{ \begin{array}{ll}  -3 + 0.5 (\beta + 3) s  & \nu_i < \nu < \nu_c \\
                  -2 + 0.5 (\beta + 1) s  & \nu_i,\nu_c < \nu  \end{array} \right.   \;.
\label{amin}
\end{equation}
Hence, for a homogeneous medium ($s=0$), the light-curve from a structured outflow could rise 
($\alpha_{min} < 0$).
Evidently, the structured outflow model can be at work only if the above decay is slower
than that observed, the condition $\alpha > \alpha_{min}$ leading to a constraint on the
CBM structure:
\begin{equation}
  s < s_{max} \equiv 2 \cdot \left\{ \begin{array}{ll} 
          \frac{\ds \alpha + 3}{\ds \beta + 3} & \nu_i < \nu < \nu_c \\
          \frac{\ds \alpha + 2}{\ds \beta + 1} & \nu_i,\nu_c < \nu   \end{array} \right. \;.
\label{smax}
\end{equation}

 Equations (\ref{SOa}) and (\ref{SOc}) give the outflow structural parameter which accommodates 
the observed light-curve index $\alpha$ and spectral slope $\beta$:
\begin{equation}
  q = \left\{ \begin{array}{ll}  
        \frac{\ds (4-s) (6\beta-4\alpha+3)+5s-12 }{\ds 2\alpha+6-(\beta+3)s } & \nu_i < \nu < \nu_c \\
        \frac{\ds (4-s) (6\beta-4\alpha  )+2s-8  }{\ds 2\alpha+4-(\beta+1)s } & \nu_i,\nu_c < \nu  
       \end{array} \right. \;.
\label{q}
\end{equation}

\subsection{Energy Injection}
\label{EI}

 Another process which can reduce the afterglow dimming rate is the injection of energy in the 
blast-wave (Paczy\'nski 1998, Rees \& M\'esz\'aros 1998) by means of some ejecta which were 
ejected later than the GRB ejecta (a long-lived engine) or at the same time but with a smaller 
Lorentz factor, thus reaching the decelerating GRB ejecta during the afterglow phase (a short-lived
engine). A delayed injection of energy into the afterglow can be due to the absorption of the 
dipole electromagnetic radiation emitted by a millisecond pulsar (Dai \& Lu 1998, Zhang \& 
M\'esz\'aros 2001) if such a pulsar was formed.

 The addition of energy in the blast-wave mitigates its deceleration and, implicitly, the afterglow 
decay rate. Rees \& M\'esz\'aros (1998) have derived the decay index $\alpha$ for an energy 
injection that is a power-law in the ejecta Lorentz factor. The expressions for the index $\alpha$ 
for an energy injection which is a power-law in the observer time,
\begin{equation}
  E_i (< t) \propto t^e \;,
\label{Ei}
\end{equation}
are given in \eqs (23), (24), and (30) of Panaitescu \& Kumar (2004). From those equations, 
it follows that energy injection reduces the light-curve decay indices given in equations 
(\ref{Sa})--(\ref{Jc}) by
\begin{equation}
 (S-EI)\; \Delta \alpha = e \cdot \left\{ \begin{array}{ll} 
               \frac{\ds 1}{\ds 2}(\beta+2) - \frac{\ds s}{\ds 8-2s}  & \nu_i < \nu < \nu_c  \\
               \frac{\ds 1}{\ds 2}(\beta+1)             & \nu_i,\nu_c < \nu \end{array} \right.
\label{Se}
\end{equation} 
\begin{equation}
 (j-EI)\; \Delta \alpha = e \cdot \left\{ \begin{array}{ll} 
        \frac{\ds 1}{\ds 2}(\beta+2) + \frac{\ds 2-s}{\ds 8-2s} & \nu_i < \nu < \nu_c  \\
        \frac{\ds 1}{\ds 2}(\beta+1) + \frac{\ds 1}{\ds 4-s}    & \nu_i,\nu_c < \nu \end{array} \right.
\end{equation} 
\begin{equation}
 (J-EI)\; \Delta \alpha = \frac{2}{3} e \cdot \left\{ \begin{array}{ll} 
               \beta+2 & \nu_i < \nu < \nu_c \\ \beta+1 & \nu_i,\nu_c < \nu \end{array} \right.
\label{Je}
\end{equation} 
for the adiabatic \S, \j, and \J models. 

 Lastly, all the decay indices given in the above equations were derived assuming that the 
microphysical parameters which determine the spectral characteristics ($\nu_i$, $\nu_c$, $F_p$) 
and the continuum slope ($\beta$), \ie the parameters for the typical post-shock electron 
energy \& magnetic field strength\footnotemark and the power-law index $p$ of the electron 
distribution with energy, are constant.  This possibility is not investigated in this work.
\footnotetext{ Yost \etal (2003) have shown that a decrease of the parameter for the magnetic 
               field energy slower than $t^{-1/2}$ or an increase slower than $t^{3/4}$ are
               allowed for several afterglows}

\section{Models for Swift $X$-ray Afterglows}

 As described in the Introduction, the Swift $X$-ray afterglows exhibit three phases:
the $1^{st}$ phase, lasting until $t_F \sim 300$ s, is characterized by a sharp decay,
the $2^{nd}$ phase, lasting until $t_S \sim 10^{3.5}-10^5$ s, is marked by a much slower fall-off, 
while in the $3^{rd}$ phase, the $X$-ray light-curve displays a faster decay. The light-curve
decay indices $\alpha$ and the spectral slopes $\beta$ are listed for each phase in table 1.
The closure relations between $\alpha$ and $\beta$ presented in section \ref{theory} provide 
either a criterion for distinguishing among the various models that can accommodate the 
observed afterglow properties or allow the determination of the CBM structure. Since $s < 3$ 
is required for a decelerating blast-wave, this also serves as a test of the various models. 

 Table 1 lists the models for which the closure relations given in section \ref{theory}
between the light-curve decay index $\alpha$ and spectral slope $\beta$ are satisfied 
within $1\sigma$, for each afterglow decay phase. To find a model for the entire afterglow,
these piece-wise models must now be put together in a sequence that makes sense and is not
contrived. The criteria by which we construct a model for the entire $X$-ray afterglow are: \\
{\bf i)} models relying on coincidences to accommodate two adjacent $X$-ray phases are excluded,
      \ie only one factor (cooling frequency passage, change of CBM structure, region of 
      non-monotonic variation in the energy per solid angle becoming visible, beginning/cessation 
      of energy injection) at a time is employed to explain a variation of the $X$-ray decay index, \\
{\bf ii)} radiative outflows can evolve into adiabatic ones, but not the other way around, \\
{\bf iii)} any of the three dynamical models (\S, \j, \J) can be followed by the same model,
      but only the \S model can be followed by the \j and \J models, allowing for a collimated
      outflow, spreading or non-spreading, whose edge becomes visible to the observer, \\
{\bf iv)} the evolution of the cooling frequency $\nu_c$ required to join two models at $t_F$
     or $t_S$ must be compatible with the CBM structural index $s$, \ie $\nu_c$ can increase
     only if $s < 4/3$ and can decrease only if $s > 4/3$ (modulo the effect of a decreasing
     Compton parameter when electron cooling is dominated by inverse Compton scatterings).

\begin{table*}
 \caption{ Models that accommodate the light-curve power-law decay index ($F_x \propto t^{-\alpha}$)
           and continuum power-law slope ($F_\nu \propto \nu^{-\beta}$) measured by Swift for the 
           early ($\sim 10^2$ s), mid ($10^3-10^4$ s), and late ($\sim 10^5$ s) $X$-ray afterglow 
           emission }
\label{T1}
\begin{tabular}{lcclcclccl}
  \hline \hline 
      & \multicolumn{3}{l}{\h $1^{st}$ phase: $t< t_F$ -- Steep Decay       } & 
        \multicolumn{3}{l}{   $2^{nd}$ phase: $t_F < t < t_S$ -- Slow Decay } &
        \multicolumn{3}{l}{   $3^{rd}$ phase: $t_S < t$ -- Fast Decay       }   \\
    GRB        & \h $\alpha_1$   & \h  $\beta_1$    &    Model          & 
                    $\alpha_2$   & \h  $\beta_2$    &    Model          & 
                    $\alpha_3$   & \h  $\beta_3$    &    Model               \\ \hline 
    050126$^a$ & \h 2.7\p.2 & \h 1.3\p.2 &    Sa,ja,Rjc,Jc            & 
                    0.6\p.1 &$\simeq 1.3$&    none                    &   
                            &            &                                 \\
    050128$^b$ &            &            &                            &
                    0.3\p.1 & \h 0.6\p.1 &    Sc (jc,RSc,Rjc)         &
                    1.3\p.2 & \h 0.8\p.1 &    Sa,jc,Jc (RSc,Rjc)           \\  
    050219A$^c$& \h 3.2\p.2 & \h 1.1\p.2 &    Sa,ja,Ja                &
                    0.8\p.1 &$\simeq 1.1$&    Sc (RSc,Rjc)            & 
                            &            &                                 \\
    050315$^a$ & \h 3.3\p.2 & \h 1.3\p.2 &    Sa,ja                   &
                    0.7\p.1 & \h 0.9\p.2 &    Sc,jc (RSc,Rjc)         &
                    1.6-2.6 & \h 1.0\p.1 &    Sa,ja,jc,Ja,Jc (RSc,Rjc)     \\
    050318$^a$ &            &            &                            &
                    1.0\p.1 & \h 1.0\p.2 &    Sa,Sc,jc,RSc (Rjc)      &
                    1.4\p.3 & \h 1.2\p.3 &    Sa,ja,jc,Jc (RSc,Rjc)        \\
    050319$^a$ & \h 3.0\p.2$^d$ & \h 1.9\p.2 &  Sa,jc,RSc,Rjc         &
                    0.5\p.1           & \h 0.8\p.1 &  Sc,jc,RSc,Rjc   &
                    1.2\p.1           & \h 0.7\p.2 &  Sa,jc,Jc (Rjc)       \\
    050401$^a$ &            &            &                            &
                    0.5\p.1 & \h 1.1\p.1 &    none                    &
                    1.6\p.1 & \h 1.1\p.1 &    Sa,jc (RSc,Rjc)              \\
    050408$^a$ &            &            &                            &
                    0.7\p.1 & \h 1.1\p.2 &    none                    &
                    1.5\p.2 &$\simeq 1.1$&    Sa,jc (RSc,Rjc)              \\
    050505$^a$ &            &            &                            &
                    0.7\p.2 & \h 1.0\p.1 &    Sc,jc (RSc,Rjc)         &
                    2.5\p.4 & \h 0.9\p.1 &    Sa,ja,Ja (Rjc)               \\
  \hline \hline 
\end{tabular}
\hspace*{5mm}
\begin{minipage}{180mm}
  Refs. for $\alpha$ and $\beta$ -- $^a$ Chincarini \etal (2005), $^b$ Campana \etal (2005),
                                    $^c$ Tagliaferri \etal (2005) \\
   $^d$ for $t=0$ at the beginning of the second GRB peak (Barthelmy \etal 2005) \\
  Model coding -- \parbox[t]{150mm}{
                  {\bf S}: spherical outflow, {\bf j}: non-spreading jet,
                  {\bf J}: sideways spreading jet, {\bf R}: radiative afterglow \\
                  (for models given in parentheses the outflow is less likely to be radiative
                  at $\sim 1$ day after the burst, or require a wind with a radial profile
                  close to $r^{-3}$, for which the analytical results given in section 
                  \ref{theory} are only approximative) \\
                  {\bf a}: $\nu_x < \nu_c$ ($X$-ray emitting electrons are cooling adiabatically) \\
                  {\bf c}: $\nu_c < \nu_x$ ($X$-ray emitting electrons are cooling radiatively) 
                   }
\end{minipage}
\end{table*}

 A structured outflow or energy injection are invoked only when the $X$-ray decay during
the $2^{nd}$ phase ($t_F < t < t_S$) is too slow to be explained by the \S, \j, and \J models,
for the CBM structured required to accommodate the $X$-ray decay preceding ($t < t_F$) or
following ($t > t_S$) this phase: \\
{\bf v)} for the structured outflow model, a working condition is that the slowest decay that 
      it would yield (\eq [\ref{amin}]), given the CBM structure which explains the $X$-ray 
      emission at $t < t_F$ or at $t > t_S$, is slower than that observed. A structured outflow 
      for the $2^{nd}$ afterglow phase cannot be preceded by the \J model, as the existence 
      of an outflow outside the jet would prevent its lateral spreading,  \\
{\bf vi)} for the energy injection model, we determine from equations (\ref{Se})--(\ref{Je}) 
      the index $e$ (\eq [\ref{Ei}]) which reconciles the slow $X$-ray decay with the spectral 
      slope, for the model (\S, \j, or \J) and CBM structure which accommodates the $X$-ray 
      emission before ($t < t_F$) or after ($t > t_S$) the energy injection episode. 
      The ratio $E_I/E_0$ of the total injected energy $E_I$ to the energy $E_0$ existing 
      in the blast-wave prior to the energy injection episode is ($t_{off}/t_{on})^e$ where 
      $t_{on}$ is the light-curve flattening time $t_F$ or the epoch of the first measurement 
      (if no flattening was observed) and $t_{off}$ is the light-curve steepening time $t_S$ or 
      the epoch of the last measurement (if no steepening was observed). Then a test of the 
      energy injection model can be done if it is assumed that the pre-injection energy $E_0$ 
      is comparable to the GRB 15--350 keV output. For this test, we calculate the injected 
      energy $E_I$ and require for the \S model that the outflow kinetic energy contained 
      within $10^{\rm o}$ (which is sufficiently wide to resemble a spherical outflow until 
      1 day after the burst) does not exceed $10^{53}$ ergs. However, this test is based
      on the assumption that the ejecta producing the burst and the beginning of the $X$-ray
      afterglow emission are the same. This does not have to be the case, as it is possible 
      that the GRB emission arises from internals shocks in the entire outflow and yet only 
      its leading edge drives the forward shock and radiates at the beginning of the $X$-ray 
      afterglow.

 We allow a variable index $s$ as we do not know what are the properties of the winds expelled 
by massive stars in the last 1,000 years before they explode, hence we do not know what is the 
density structure of the CBM within the first parsec, where the afterglow emission is produced. 
Variations in the GRB progenitor's mass-loss rate and wind speed could lead to a CBM with a 
structure different than the $r^{-2}$ profile expected for a constant speed, constant mass-loss 
rate, and to interactions between winds that could form shells of higher density.

 By applying the above criteria, we arrive at the models given in Table 2. Below we discuss
in some detail the 9 Swift $X$-ray afterglows and the models that accommodates them.

\begin{table*}
 \caption{ Possible models for the afterglow X-ray phases of Table 1  }
\label{T2}
\begin{tabular}{llclccclcc}
  \hline \hline
      & \multicolumn{2}{l}{$t < t_F$ -- Steep Decay} & 
        \multicolumn{4}{l}{$t_F < t < t_S$ -- Slow Decay} &
        \multicolumn{2}{l}{$t_S < t$ -- Fast Decay} &             \\
  GRB   & Model &    s    & Model &    s    &    e   &   q    &  Model  &    s    &    p     \\
        &       &   (1)   &       &   (1)   &   (2)  &  (3)   &         &   (1)   &   (4)    \\
  \hline
 050126 &  ja   & $<2.5$  &  EI   &         &1.0--1.4&        &         &         &  3.5\p.4 \\
        &  ja   & $<2.5$  &  SO   & $<1.7$  &        & $>0.7$ &         &         &  3.5\p.4 \\
        &  Rjc  & $ <3 $  &  EI   &         &1.0--1.9&         &         &         &  2.5\p.4 \\ 
        &  Rjc  & $ <3 $  &  SO   & $<2.3$  &        & $>1.4$ &         &         &  2.5\p.4 \\ 
        &  Jc   & $ <3 $  &  EI   &         &   1.4  &        &         &         &  2.5\p.4 \\ 
 \hline
 050128 &       &         &  Sc   & $ < 3 $ &        &        &  jc(Jc) &$<2.2(3)$&  1.2\p.1 \\
        &       &         &  jc   &   3     &        &        &    jc   & $ < 0 $ &  1.2\p.1 \\
 \hline         
 050219A&  ja   & 2.4--3  &  EI   &         &1.7--2.3&        &         &         &  3.2\p.4 \\
        &  Ja   & $ < 3 $ &  EI   &         &   1.2  &        &         &         &  3.2\p.4 \\
 \hline
 050315 &LA--GRB&         &  Sc   & $ < 3 $ &        &        &  jc(Jc) &$<2.1(3)$&  1.9\p.1 \\ 
 \hline
 050318 &       &         &  Sc   & $ < 3 $ &        &        &  jc(Jc) &$<1.9(3)$&  2.1\p.1 \\
        &       &         &  Sa   &  $<-3$  &        &        &    Sa   &    2    &  3.0\p.3 \\
        &       &         &  jc   &  2.6--3 &        &        &  jc & $\rightarrow-\infty$ & 2.1\p.1 \\
 \hline
 050319 &LA--GRB&         &  Sc   & $ < 3 $ &        &        &  jc(Jc) &$<2.1(3)$&  1.5\p.2 \\
        &LA--GRB&         &  jc   &    3    &        &        &    jc   &    0    &  1.5\p.2 \\
 \hline
 050401 &       &         &  EI   &         &   0.7  &        &    Sa   & -1--0.6 &  3.1\p.1 \\
        &       &         &  SO   & $<1.7$  &        &1.7--3.5&    Sa   & -1--0.6 &  3.1\p.1 \\
        &       &         &  SO   & $<2.4$  &        & $>1.1$ &    RSc  & $ < 3 $ &  2.1\p.1 \\   
        &       &         &  EI   &         &   0.7  &        &    jc   & 1.6--2.4&  2.1\p.1 \\
        &       &         &  SO   & $<2.4$  &        & $>3.2$ &    jc   & 1.6--2.4&  2.1\p.1 \\
        &       &         &  EI   &         &   0.6  &        &    Rjc  & 2.7--2.9&  2.1\p.1 \\
 \hline
 050408 &       &         &  EI   &         &0.4--0.6&        &    Sa   & $<1.1$  &  3.3\p.4 \\
        &       &         &  SO   & $<1.8$  &        & $>0.5$ &    Sa   & $<1.1$  &  3.3\p.4 \\
        &       &         &  EI   &         &0.4--0.6&        &    jc   & 0.7--3  &  2.3\p.4 \\
        &       &         &  SO   & $<2.5$  &        & $>0.8$ &    jc   & 0.7--3  &  2.3\p.4 \\
 \hline
 050505 &       &         &  EI   &         & $>0.9$ &        &   ja/Ja & $ < 3 $ &  2.8\p.2 \\
        &       &         &  SO   & $<1.9$  &        & $>0.9$ &   ja    & -2--3   &  2.8\p.2 \\
 \hline \hline
\end{tabular}
\hspace*{0mm}
\begin{minipage}{170mm}
  Model coding -- \parbox[t]{120mm}{ EI: energy injection; SO: structured outflow;
                                     LA--GRB: large-angle ($\theta > 1/\Gamma$) GRB emission } \\
 (1): exponent of radial density profile (\eq [\ref{n}]); 
       for the SO model, the upper limit on $s$ is that resulting from equation (\ref{smax}) \\ 
 (2): exponent of the energy injection law (\eq [\ref{Ei}]) obtained from \eqs (\ref{Se})--(\ref{Je}) 
       for the index $s$ required at $t < t_F$ or at $t > t_S$ \\
 (3): exponent of the angular distribution of the energy per solid angle (\eq [\ref{Etheta}]) 
       obtained from equation (\ref{q}) for the index $s$ required at $t < t_F$ or at $t > t_S$ \\
 (4): exponent of the power-law distribution of electrons with energy (\eq [\ref{N}]);
          this value is for all $X$-ray phases except LA--GRB 
\end{minipage}
\end{table*}

{\bf 050126.}
 The XRT light-curve of this afterglow exhibits a steep fall-off until $t_F \simg 300$ s, 
followed by a slower decay. Tagliaferri \etal (2005) have shown that extrapolation of the 
15--350 keV BAT emission to the 0.2--10 keV XRT band is dimmer at 100 s than the observed 
XRT flux. Furthermore, the XRT spectrum ($\beta_1 = 1.26 \pm 0.22$) during the $1^{st}$
phase is softer than the BAT spectrum ($\beta_\gamma = 0.32 \pm 0.18$), hence the early 
$X$-ray afterglow is not the large-angle GRB emission and must be attributed to the forward 
shock. If there is no spectral evolution ($\beta_2=\beta_1$) across $t_F$, as indicated
by Tagliaferri \etal (2005), then the slow $X$-ray decay of the $2^{nd}$ phase cannot be 
explained by a change in the structure of the CBM medium for any of the models (\Sa, \ja, 
\Rjc, \Jc) which accommodate the $1^{st}$ phase. Conversely, if the CBM structure does 
not change across $t_F$, then the slower decay at the $2^{nd}$ phase requires a substantial 
hardening of the spectrum, corresponding to a rising one ($\beta_2 < 0$) for the models 
\Sa, \ja, and \Rjc, or one with $\beta_2 = 0.20 \pm 0.25$ for the \Jc model, both of which
are inconsistent with the XRT observations. Furthermore, for the possible models for the 
$1^{st}$ afterglow phase, the passage of the cooling frequency through the $X$-ray band can 
only steepen the afterglow decay. Hence, the most plausible models that can explain the 
flattening $X$-ray light-curve of 050126 require energy injection or a structured outflow.
The \Sa model with either energy injection or a structured outflow does not satisfy conditions
$v)$ and $vi)$ above.

{\bf 050128.}
 Although XRT observations started at 100 s after the burst, a steep early decay has not 
been observed (Campana \etal 2005). Its decay steepens at $t_S \simg 10^3$ s, without a 
spectral evolution. Of the many possible combinations of models for the $2^{nd}$ and $3^{rd}$ 
phases, the most plausible is that of a collimated outflow (\jc and \Jc models), leading 
to a steepening of the $X$-ray decay when the boundary of the jet becomes visible. 
Another possibility is that of non-spreading jet (\jc model) which transits from a $r^{-3}$ 
wind into a region of increasing density at $t_S$. We note that all these models require a 
rather hard electron distribution, with $p \siml 1.3$.

{\bf 050219A.}
 The features of this afterglow are similar to those of 050126. It exhibits a fast fall-off 
until $t_F \sim 300$ s, followed by a slower decay. The extrapolation of the 15--350 keV BAT 
emission to the 0.2--10 keV XRT band underpredicts the observed flux at 100 s (Tagliaferri 
\etal 2005) and the $X$-ray spectral slope ($\beta_1 = 1.1 \pm 0.2$) is much softer than 
that of the burst ($\beta_\gamma = -0.75 \pm 0.30$), hence the rapid, early fall-off of the 
050219A $X$-ray afterglow is not the GRB large-angle emission. Just as for the afterglow 
050126, a change in the CBM structure cannot explain the $X$-ray light-curve flattening. 
If the CBM structure is considered unchanged across $t_F$, then the slowing of the $X$-ray 
decay would require a rising spectrum ($\beta_2 < 0$) for the $2^{nd}$ phase, which is inconsistent
with the XRT observations. Because all models for the $1^{st}$ afterglow phase require that 
the cooling frequency is above the $X$-ray domain, its passage is either impossible or it 
would steepen the light-curve decay. Consequently, the slower decay observed for the $X$-ray 
afterglow 050219A after $t_F$ requires either energy injection or a structured outflow. 
Condition $v)$ is not satisfied by either the \Sa and \ja models and a structured outflow,
while the \Sa model with energy injection requires too much energy.

{\bf 050315.}
 The $X$-ray emission exhibits a flattening at $t_F \siml 10^3$ s, accompanied by a hardening 
of the spectrum ($\beta_2-\beta_1 = -0.41 \pm 0.21$), and followed by a steeper decay after 
$t_S \simg 10^5$ s, across which there is no spectral evolution. Barthelmy \etal (2005) have 
shown that the early, steep fall-off is consistent with the large-angle GRB emission: the 
extrapolation of the 15--350 keV BAT emission to the 0.2-10 keV XRT band matches the XRT 
flux measured at 100 s, the $X$-ray spectral slope ($\beta_1 = 1.34 \pm 0.15$) is comparable 
to that of the burst ($\beta_\gamma = 1.18 \pm 0.11$), and the $X$-ray decay index ($\alpha_1 
= 3.35 \pm 0.19$) is close to the expected value ($2+\beta_\gamma = 3.18 \pm 0.11$). 
The steepening at $t_S$ can be easily understood as due to a collimated outflow (the \j or \J 
models). A radiative non-spreading jet interacting with $s \siml 3$ CBM could also accommodate 
the steepening, if the CBM is a wind, but it is less likely that the radiative phase could 
last until later than 1 day after the burst.

{\bf 050318.}
 Because XRT observations started at $\siml 1$ h after the burst, the fast decay phase may have 
been missed. A steepening of the $X$-ray light-curve decay occurs at $t_S \sim 3 \times 10^4$ s 
without a spectral evolution. This steepening can be due to seeing the boundary of a jet 
(spreading or not). There are other possible models that can accommodate the steepening, all 
involving a variation in the CBM structural index $s$. They are the \Sa outflow exiting a shell 
of a sharply increasing density and entering a $r^{-2}$ wind and \a jc outflow transiting from 
a $r^{-3}$ wind to a shell with sharply increasing density at $t_S$.

{\bf 050319.}
 This afterglow is similar to 050315, the hardening of the $X$-ray spectrum across the 
light-curve flattening, which occurs at $t_F \sim 400 s$, being stronger. Barthelmy
\etal (2005) have shown that the BAT GRB emission extrapolated to the XRT band matches the
$X$-ray flux measured at 200 s. If the origin of time for the $X$-ray emission is set at the
beginning of the second (and last) GRB pulse, then the decay index ($\alpha_1 = 3.0 \pm 0.2$) 
of the early $X$-ray emission is consistent with the expectations for the large-angle GRB 
emission ($2+\beta_\gamma = 3.13 \pm 0.28$). However, the early $X$-ray spectrum ($\beta_1 = 
1.94 \pm 0.20$) is rather soft compared to that of the burst. On the other hand, the substantial 
hardening of the $X$-ray spectrum across $t_F$, with $\beta_2 - \beta_1 = -1.15 \pm 0.23$, 
exceeds that which the passage of the cooling frequency through the observing band can produce 
($\beta_2 -\beta_1 = -0.5$), suggesting that the $X$-ray emissions during the $1^{st}$ and $2^{nd}$ 
afterglow phases arise from different mechanisms. We note that the $X$-ray light-curve for both 
phases may be explained in the structured outflow framework if we make the ad-hoc 
assumption that the spectrum of the spot emission (dominating the afterglow flux before $t_F$) 
is softer than that from the surrounding outflow (which overtakes the spot emission after $t_F$).

 The steepening of the $X$-ray light-curve at $t_S \sim 3 \times 10^4$ s can be explained
by seeing the edge of a jet (spreading or not), or with the \jc model and a CBM structure
changing from a $r^{-3}$ wind to a homogeneous medium at $t_S$. All these models require a 
hard electron distribution, with $p < 1.7$. 

{\bf 050401.}
 Although the XRT observations started 100 s after the burst, a steeply falling-off phase
was not seen. Until $t_S = 5000$ s, it exhibits a decay so slow that it cannot be explained 
without energy injection or a structured outflow. The \RSc model with energy injection requires
too much energy, while the structured outflow does not satisfy condition $v)$ for the \Rjc model. 
Then the steepening of the $X$-ray light-curve at $t_S$ can be understood either as resulting 
from the cessation of the energy injection or from seeing the outflow boundary. 
In the latter case, the light-curve decay should be steeper than for the \S model and slower 
than for the \j model. That the steeper decay after $t_S$ can be accommodated by either the 
\S and \j models (Table 1) supports a structured outflow as the source of the $X$-ray light-curve
steepening.

{\bf 050408.}
 This afterglow is very similar to 050401, except that the steepening occurs later, at $t_S \sim 
10^5$ s. Its light-curve decay before $t_S$ is also too slow and requires an energy injection 
episode or a structured outflow. The $X$-ray spectral slope after $t_S$ is not known, but if we 
assume that there is no spectral evolution across $t_S$ (as is the case for all other afterglows), 
then the light-curve decay index and spectral slope measured after $t_S$ can be accommodated by 
the \Sa and \jc models. The \RSc and \Rjc models are also allowed, though it is unlikely that 
the radiative phase lasts until days after the burst.

{\bf 050505.}
 This afterglow is similar to 050318, its $X$-ray light-curve steepening at $t_S = 4 \times
10^4$ s without a spectral evolution. However, in contrast to 050318, the spectral slopes
and decay indices before and after $t_S$ cannot be reconciled within any model other than
\Rjc, even if we allow for a varying CBM structure. Besides that the radiative phase is 
unlikely to last until 1 day after the burst, the \Rjc model requires a $r^{-3}$ CBM profile, 
for which the closure equations given in section \ref{radiative} are not accurate. Hence,
it seems more plausible that the slow decay of this afterglow before $t_S$ is due to energy 
injection or a structured outflow. The \Sa model fails to satisfy conditions $v)$ and $vi)$
for these two case. As for the afterglows 050401 and 050408, the steepening of the $X$-ray 
light-curve could then be attributed to the end of the energy injection or to the outflow 
axis becoming visible to the observer.

\section{Discussion}
 
 As shown in Table 1, the three decay phases of the Swift $X$-ray afterglows can be understood
in the following way:
$i)$ the hardening of the 0.2--10 keV spectrum of the $X$-ray afterglows 050315 and 050319
  from $t < 400$ s (when a fast decaying $X$-ray emission is observed) to $t > 400$ s (when
  the $X$-ray light-curve exhibits a slow decay) indicates that the early, fast-falling off
  $X$-ray emission arises from a different mechanism than the rest of the afterglow. 
  Barthelmy \etal (2005) have argued that this mechanism is the same as for the GRB emission.
  The results of Tagliaferri \etal (2005) suggest that this explanation does not work well
  for the afterglows 050126 and 05018. In these cases, the steep $X$-ray decay require a very 
  narrow outflow whose edge is seen as early as 100 s after the burst, the following, slower 
  decay phase being explained by energy injection in the blast-wave or by a rather contrived
  (see below) angular structure of the blast-wave, \\
$ii)$ the $X$-ray decay measured until 1h, 0.3d, and 0.5d for the afterglows 050401, 050408, 
  and 050505, respectively, is too slow to be explained by the simplest blast-wave model. 
  Such a slow decay can be produced by an outflow endowed with angular structure or by a 
  continuous injection of energy in the forward shock, \\
$iii)$ the steepening of the $X$-ray decay observed at 1h--2d for the afterglows 050128,
  050315, 050318, and 050319, \ie at a time comparable to that of the steepening of the 
  optical decay of many pre-Swift afterglows, can be explained by seeing the edge of a jet.
  For the remaining 5 afterglows, whose pre-break $X$-ray decay requires energy injection,
  the steepening can be attributed to the cessation of injection. 

 In the large-angle GRB emission model for the early, fast falling-off phase, the $X$-ray 
emission arises from the same mechanism as the GRB itself, but arrives at observer later 
because it comes from the shocked gas moving slightly off the direction toward the observer. 
For this model to be at work, three conditions must be satisfied. 
First, the 15--350 keV GRB emission extrapolated to the 0.2--10 keV $X$-ray band, under the 
assumption that the power-law burst spectrum $F_\nu \propto \nu^{-\beta_\gamma}$ extends 
unbroken down to 0.2 keV, should match or exceed the $X$-ray flux at the first epoch of observations. 
Second, the spectral slope of the early afterglow should be the same as that of the GRB. 
Third, the $X$-ray light-curve decay index should be equal to $2+\beta_\gamma$ (Kumar \& Panaitescu 2000). 
For completeness, we present here a short derivation of this result. 
If the GR emission stops suddenly at some radius $r$ and blast-wave Lorentz factor $\Gamma$, 
then the received flux is 
  $F_\nu (t) \propto {\cal F}'_{\nu'} (\d\Sigma/\d t) {\cal D}^2$ where 
 ${\cal F}'_{\nu'} \propto \nu'^{-\beta}$ is the outflow comoving frame surface-brightness 
             at frequency $\nu' = \nu/{\cal D}$, 
 $\d\Sigma = 2\pi r^2 \theta \d\theta$ is the elementary area whose radiation is received 
             over an observer time $\d t$, 
 $\theta$ is the angle (measured from the direction toward the observer) of the fluid element
  from which radiation is received at time $t = r\theta^2/2$ (hence $\d t \propto \theta \d\theta$), 
 ${\cal D} = 2/(\Gamma \theta^2) \propto t^{-1}$ is the relativistic Doppler factor,  
 and the expressions for $t(\theta)$ and ${\cal D}(\theta)$ have been derived for 
             $\theta \gg \Gamma^{-1}$, \ie for the large-angle emission.
 The last factor ${\cal D}^2$ in the expression of $F_\nu$ accounts for the beaming of 
     radiation from a relativistic source. 
 After substitutions, one obtains that $F_\nu \propto \nu^{-\beta} t^{-2-\beta}$. 

 Barthelmy \etal (2005) found that the above three conditions for the large-angle GRB emission
as the source of the very early, fast $X$-ray decay are met for the afterglows 050315 and 050319. 
For two other afterglows, 050421 (Sakamoto \etal 2005, Godet \etal 2005) and 050713B 
(Parsons \etal 2005, Page \etal 2005), we find that their fast $X$-ray decays 
cannot be reconciled with their hard $X$-ray continua by any of the blast-wave models considered 
here, but they do satisfy the last two conditions above for the large-angle GRB emission 
interpretation. Tagliaferri \etal (2005) showed that the early $X$-ray emissions of the afterglows 
050126 and 050219A are brighter than the GRB extrapolated fluxes and softer than the burst 
emission, therefore their early $X$-ray afterglows cannot be immediately identified with the 
large-angle GRB emission. Kumar \etal (2005) discuss the conditions under which the fast $X$-ray 
decay of these last two afterglows can be reconciled with the large-angle GRB emission. 
 
 The outflow structure required to explain a light-curve flattening followed by a steepening 
must contain a bright spot (moving toward the observer) surrounded by a dimmer envelope where
the ejecta have a lower energy per solid angle $\d E/\d\Omega$, so that the emission from the
spot exhibits a fast decay after its boundary becomes visible. Further, the envelope should be
embedded in a wider outflow whose emission overtakes that of the spot when the blast-wave Lorentz 
factor has decreased sufficiently. To explain the light-curve flattening, the $\d E/\d\Omega$ 
in this wider outflow should rise away from the spot as $\theta^{1/2}$ to $\theta^{3}$, where
$\theta$ is the angle measured from the outflow's symmetry axis. Finally, to explain the light-curve
steepening, the $\d E/\d\Omega$ should stop increasing with angle (for the \S model) or peak 
and then decrease (for the \j and \J models).
The decrease could be gradual, with the $X$-ray light-curve steepening occurring when the fluid 
at the peak of $\d E/\d\Omega$ becomes visible and the outer outflow contributing to the post-break 
emission (this is the light-curve break mechanism proposed by Rossi, Lazzati \& Rees 2002). 
If the decrease of $\d E/\d\Omega$ is sharp, then the post-break $X$-ray light-curve decay will 
be faster, particularly if the outflow undergoes lateral spreading (this is the light-curve 
break mechanism proposed by Rhoads 1999). 

 In the energy injection model, the forward shock energy increases due to some relativistic ejecta 
which catch up with the decelerating blast-wave (Rees \& M\'esz\'aros 1998). The energy injection 
reduces the blast-wave deceleration rate and mitigates the decay of the afterglow emission. 
In principle, during the slow $X$-ray decay phase, there could be an energy injection for all 
afterglows considered here; Table 2 lists only the cases when it is required. As shown in Table 2, 
we find that, to explain the slow phase of the $X$-ray afterglow decay, the blast-wave energy 
should increase with observer time faster than $t^{0.5}$ and slower than $t^{1.5}$. 
The arrival of new ejecta at the forward shock could be due either to the spread in the Lorentz 
factor of ejecta released simultaneously (short-lived engine) or to a long-lived GRB engine, 
releasing a relativistic outflow for a source-frame duration comparable to the observer-frame 
duration of the slow $X$-ray decay phase. In the former case, the ejecta--forward-shock
contrast Lorentz factor is $[(4-s)/(1+e)]^{1/2} \siml 2$ (Panaitescu \& Kumar 2004), where $e$ is 
defined by equation (\ref{Ei}), while in the latter case the Lorentz factor ratio can be much larger. 
Consequently, we expect that, for a short-lived engine, the reverse shock propagating in the
incoming ejecta is only mildly relativistic and radiates mostly at radio frequencies while
for a long-lived engine the reverse shock could be very relativistic and radiate in the 
infrared-optical.

 As mentioned above, the early, fast decay of the $X$-ray afterglows 050126 and 050219A cannot be 
readily identified with the large-angle GRB emission and could be attributed to the forward shock.
The fast $X$-ray decay displayed by these afterglows at $\sim 400$ s can then be explained only 
within the structured outflow and energy injection models.
In addition, two other afterglows, not considered in this work, 050712 (Grupe \etal 2005) and 
050713A (Morris \etal 2005), exhibit an $X$-ray decay which is too slow and incompatible with the 
reported $X$-ray spectral slope, both requiring either an energy injection or a structured outflow.

 For the afterglows 050128, 050318, and 050319, we find that a change in the circumburst density 
profile provides an alternate model to structured outflows and jets for the steepening of the 
$X$-ray decay observed by XRT after 0.1 d. For all three afterglows, the changing external density 
corresponds to a transition from a $r^{-3}$ wind to a region of uniform or increasing density.
The $r^{-3}$ density structure requires a time-varying mass-loss rate and/or speed of the wind of 
the massive star GRB progenitor, while the uniform or increasing density shell could result from 
the internal interactions in a variable wind (\eg Ramirez-Ruiz \etal 2005). The self-similar solutions 
of Chevalier \& Imamura (1983) for wind-wind interactions indicate that a uniform shell results 
from a substantial decrease of the star's mass-loss rate accompanied by a large increase in the wind 
speed. These major changes in the wind properties would have to occur $\sim 1,000$ yrs before the 
GRB explosion, if the radius where the $r^{-3}$ circumburst density profile terminates is that of 
the forward shock at 0.1--1 d.
 
 To answer the question of how can we distinguish between the above three models (energy 
injection, structured outflow, non-monotonic circumburst density) for flattenings and steepenings 
of the afterglow light-curve, we note that, if the cooling frequency located between the optical 
and $X$-ray domains, each of those models yields a specific difference $\Delta \alpha_c - 
\Delta \alpha_o$ between the changes $\Delta \alpha_c$ and $\Delta \alpha_o$ of the $X$-ray 
and optical light-curve decay indices. Therefore, to discriminate among the few possible models 
discussed here, it is very important to monitor the optical afterglow emission over a wide range 
of times, from minutes to days after the burst.

\section{Conclusions}

 The most often encountered feature resulting from the analysis of the nine $X$-ray afterglows
analyzed in this work is the existence of a substantial energy injection in the blast-wave at hours 
to 1 d after the burst. This energy injection is necessary to reconcile the slowness of the $X$-ray 
decay with the hardness of the $X$-ray continuum for five out of nine afterglows and is possible 
for the other four as well. The injection should increase the blast-wave energy as $t^{1.0\pm0.5}$, 
$t$ being the observer time, leading to a shock energy which is eventually larger by a factor 
10--1,000 than that at the beginning of the slow $X$-ray decay phase. 

 The exponent $e$ of power-law energy injection identified in Table \ref{T2} is generally 
inconsistent with that expected for the absorption of the dipole radiation from a millisecond 
pulsar (see also Zhang \etal 2005). In this case, an important energy input in the blast-wave 
is obtained only for the first few thousand seconds, when the pulsar electromagnetic luminosity 
is constant (Zhang \& M\'esz\'aros 2001), which leads to $e=1$. Hence the energy injection must 
be mediated by the arrival of new ejecta at the forward-shock. 

 If the GRB engine were so long-lived that only a small fraction of the total outflow energy 
yielded the burst emission, then such a large energy injection would imply an even higher GRB 
efficiency than previously inferred (above 30 percent -- Lloyd-Ronning \& Zhang 2004) from afterglow 
energetics and would pose a serious issue for the GRB mechanism (Nousek \etal 2005). However, we 
cannot yet tell if the energy injection lasts for
hours because the central engine is long-lived or because there is a sufficiently wide distribution 
of the initial Lorentz factor of the ejecta expelled by a short-lived engine. In the latter case,
the entire outflow could be emitting both the GRB and afterglow emission and the GRB efficiency 
remains unchanged. 

 If the fast decaying $X$-ray emission preceding the slow $X$-ray fall-off arises from the forward 
shock as well, then energy injection must start at the end of the steep decay phase, \ie it must
be a well defined episode. However, in this scenario the outflow must be very tightly collimated
to yield a fast decay at only 100 seconds after the burst. The blast-wave Lorentz factor is
$\Gamma (t=100s) \simeq 100 (E_{53}/n_0)^{1/8}$ ($E_{53}$ being the shock energy in $10^{53}$ ergs 
and $n_0$ the circumburst medium particle density in ${\rm cm}^{-3}$), thus the jet opening must
be less than $\Gamma^{-1} = 0.5^{\rm o}$. Alternatively, the early fast falling-off $X$-ray emission 
could be the large-angle emission from the burst phase, overshining the forward shock emission. 
This interpretation is favoured by Barthelmy \etal (2005) and Hill \etal (2005) for the afterglows
050117, 050315, 0509319, although for other afterglows (\eg 050126 and 050219A -- Tagliaferri \etal 
2005) the spectral properties of the burst and $X$-ray afterglow emissions are not readily consistent 
with each other. 

 For a structured outflow to explain all the three decay phases of the Swift $X$-ray afterglows,
the distribution of the ejecta kinetic energy with angle must be non-monotonic. This is somewhat 
contrived and inconsistent with the results obtained by MacFadyen, Woosley \& Heger (2001) from 
simulations of jet propagation in the collapsar model. Thus structured outflows do not appear 
to provide a natural explanation for the features of the $X$-ray afterglow light-curves. 

 We note that the inferred indices of the power-law electron distribution with energy, which are 
given in Table 2, range from 1.3 to 2.8. One would have to ignore 4 of these 9 Swift afterglows 
to obtain a unique electron index, $p = 2.1 \pm 0.1$. This is a puzzling feature of relativistic 
shocks in GRB afterglows: the shock-accelerated electrons do not have a universal distribution with 
energy, a fact which is also proven by the wide spread of the high-energy spectral slopes of BATSE 
bursts ($\Delta \beta \simeq 2.0$ in fig. 9 of Preece \etal 2000), which is equal to $p/2$ or $(p-1)/2$, 
and the by wide range of the optical post-break decay indices of the BeppoSAX afterglows ($\Delta 
\alpha \simeq 1.6$ in fig. 3 of Stanek \etal 2001 and in fig.2 of Zeh, Klose \& Kann 2005), which 
is equal to $p$. However, from the $X$-ray spectral slope of 15 BeppoSAX afterglows, De Pasquale 
\etal (2005) conclude that the electron index $p$ has an universal value of $p = 2.4 \pm 0.2$ 
(see their fig. 3). 
  
\vspace*{4mm} \noindent
{\bf Acknowledgments.} We are indebted to Pawan Kumar for his comments on this work.

\end{document}